\documentclass{aa}

\usepackage[varg]{txfonts}

\usepackage{hyperref}
\usepackage{natbib}
\bibpunct{(}{)}{;}{a}{}{,} 

\def\msun{{ ~M}_{\odot}}

\newcommand{\Ledd}{\ensuremath{\,L_{\rm Edd}}}
\newcommand{\delj}{\ensuremath{\,\delta_{\mathcal{J}}}}
\usepackage{color}

\newcommand{\jp}[1]{{\color{red} {#1}}} %

\begin{document}

   \title{{The stress at the ISCO of black-hole accretion discs is not a free parameter} }
   

   \author{Jean-Pierre Lasota\inst{1,2}
   \and 
   Marek Abramowicz\inst{3,1,4}  
   }

   \institute{Nicolaus Copernicus Astronomical Center, Polish Academy of Sciences,
          ul. Bartycka 18, 00-716 Warsaw, Poland                       
         \and 
            Institut d'Astrophysique de Paris, CNRS et Sorbonne Universit\'e,
           UMR 7095, 98bis Bd Arago, 75014 Paris, France
           \and
         Research Centre for Computational Physics and Data Processing; 
            Institute of Physics, Silesian University in Opava, Bezru\v{c}ovo n\'am. 13, Opava 746 01, Czech Republic 
            \and
             Department of Physics, G{\"o}teborg University, SE-412-96 G{\"o}teborg, Sweden
             }

             \offprints{J.-P. Lasota, \email{lasota@iap.fr}\\ }

   \date{Received \today; accepted ...}

 
  \abstract{Radiation from the ``plunging region'' (the space between the innermost stable circular orbit (ISCO) and the black-hole (BH) surface) of an accretion flow onto a BH is supposed to add a small but significant contribution to the X-ray spectra of X-ray binary systems. The plunging region and its electromagnetic emission has been recently described by numerical and analytic calculations which lead to the conclusion that in the plunging region  radiation  is generated by the energy release through the action of a ``leftover'' stress in the vicinity of the ISCO and that the amount of this leftover can be chosen as a free parameter of the accretion-flow description. The present article aims to demonstrate that this is not true because the stress in the whole accretion flow onto a black hole is determined by its global transonic character enforced by the space-time structure of the accreting black hole.
  We use the slim-disc approximation (SDA) to illustrate this point. In our article we compare models obtained with the SDA with results of numerical simulations and of analytical models based on the assumption that accreting matter flows along geodesics. We show that the latter cannot describe adequately the physics of astrophysical accretion onto a BH because 1. particles on geodesics cannot emit electromagnetic radiation, 2. they ignore the global transonic character of the accretion flow imposed by the presence of a stationary horizon in the BH spacetime; a presence that fixes a unique value of the angular momentum at the BH surface for a solution to exist. Therefore the fact that geodesics-based models reproduce the trans-ISCO behaviour of GRRMHD simulations proves the physical reality of neither. We show also that the claimed detection of plunging-region emission in the spectrum of an X-ray binary is unsubstantiated. 
   
}


   \titlerunning{Stress at the ISCO}

   \maketitle

   \nolinenumbers
   
%

\section{Introduction}
\label{sec.intro}

In recent years the so-called plunging region of the accretion flow onto black holes (BHs) has attracted
a lot of attention \citep{Zhu0812,Mummery0523,Mummery0424,Mummery0624,Mummery0824}. The plunging
region (``the eye of the storm'' according to \citealt{Zhu0812}) is defined as the part of the accretion 
flow located below the innermost stable circular orbit
and the main motivation behind the studies of its properties is the hope of determining the location
of the ISCO from fitting observed observed X-ray spectra with models of accretion disc spectra, which according
to a wide-spread belief would lead to the determination of the accreting BH spin. This belief is fundamentally
unsubstantiated as demonstrated by \citet{Belczynski24,Belczynski1024} and \citet{Zdziarski0224a,Zdziarski0224b} but it does not imply
that the problem of the nature of the accretion flow ``leaving the ISCO'' is not interesting in itself. In fact
it has been addressed in great detail by \citet{Abramowicz1010} (see also \citealt{Krolik0702}) but these papers 
are largely ignored by people calculating radiation from matter plunging into a BH..

This might be one of the reasons why a fundamental property of accretion of matter onto BHs is ignored, namely that
it is described by a set of differential equations with two critical points \citep{ShapiroTeukolsky,Abramowicz0988,Abramowicz0189,Abramowicz1196,Lasota0194,Afshordi0703,Sadowski0811}.
One of the critical points is located at the horizon $r=r_H$ and requires the proper radial velocity of the flow to
be equal to the speed of light $V=1$  (in what follows we assume $G=c=1$) independent
of the magnitude of the flow's four-velocity $u^i$. This just follows from the fact
that the surface of a BH is a stationary horizon, i.e. a stationary null hypersurface and thus puts inescapable 
constraints on any physically acceptable solution representing accretion onto a BH.

The second critical point at $r=r_s$ corresponds to $V=a_s$, where $a_s$ is
the adiabatic speed of sound so it is the sonic point. An accretion flow onto a BH with $a_s<1$ must pass through
the sonic point and be critically transonic since in the accretion disc $V<a_s$.
One should stress that although for $r \rightarrow \infty$ one has $V \rightarrow 0$ the requirement $V=1$ at the horizon
is not sufficient by itself to guarantee that the solution passes through the critical sonic point since this condition
is not a property of the flow but of the spacetime.

Therefore for a set of parameters describing the properties of the spacetime and the accretion flow the
set of equations describing accretion onto a BH possess a {\sl unique}, critically transonic solution.
Do the descriptions of the plunging region by \citet{Zhu0812,Mummery0523,Mummery0424,Mummery0624,Mummery0824}
correspond to such solutions? We cannot answer this
question because their authors neither address it, nor provide the tools that would allow finding an answer,
but if they do not, then they are not physically valid (conserving matter, momentum and angular momentum) descriptions
of accretion onto BH. One should observe, however, since the GRRMHD simulations by \citet{Zhu0812} basically solve
the same equations \citep{Gammie0503,Mckinney0309}\footnote{In addition they also solve Maxwell equations but this does not modify the nature of the problem.} as e.g. \citet{Abramowicz1196} the numerical solutions they obtain are obviously subject
to the same constraints.

Since the critically transonic solution describing accretion onto a BH is unique it precludes making arbitrary
assumptions about the value of some physical quantities at a given distance from the BH. Indeed, the regularity condition
at the sonic point is satisfied for only one value $\ell_0$ \citep{Abramowicz0189} of the angular momentum at the horizon, where
the torque must be, mathematically and physically equal to zero. \citet{Paczynski0400} argues that for geometrically
thin discs and low viscosity parameters $\alpha$ the ‘no torque inner boundary condition’ at ISCO is an excellent
approximation of the exact solutions. To prove this Paczy\'nski uses the equation \footnote{Paczy\'nski uses here the \citet{PW} framework but
this has no influence on the conclusions.}
\begin{equation}
\label{eq:bp}
    \frac{V}{c_s}=1 \approx \alpha \frac{H_{\rm in}}{r_{\rm in}}\frac{\ell_{\rm in}}{\ell_{\rm in} - \ell_0}
    \ \ \ \ \ \ \ {\rm at}\ \ r=r_{\rm in},
\end{equation}
describing the flow parameters at the sonic point located near the inner disc edge $r_{\rm in} \approx r_{\rm ISCO}$.
When $\alpha {H_{\rm in}}/{r_{\rm in}} \ll 1$, on has ${\ell_{\rm in} \approx \ell_0}$ implying indeed a very small torque
at $r=r_{\rm in}$. 

However, it should be stressed that for a given critical transonic solution, i.e. for a given value of $\ell_0$ (and $\alpha$), the values of $l_{\rm in}$ and $\alpha{H_{\rm in}}/{r_{\rm in}}$ are {\sl fixed} by this solution and cannot be played with contrary to what e.g. \citet{Mummery0624} are doing. This fundamental property of accretion onto BHs is the main subject of our paper.

In the next section  we will begin by reminding the basics of general-relativistic accretion onto a BH and in section \ref{sec:equations}
we will present the equations describing the slim disc approximation (SDA) equations. Section \ref{sec:glob} is devoted to the
description of the nature of the global transonic solutions describing accretion flows onto BHs. In section \ref{sec:versus}
we compare critically the global transonic solutions obtained in the SDA with solutions obtained with
GRRMHD simulations and with the geodesics models. In this section we also show that the assertion of the first detection of
plunging-region emission is unfounded.
Section \ref{sec:discussion} contains a discussion of the nature of general-relativistic
accretion onto BHs and we end the paper with conclusions in section \ref{sec:conclusions}.

Following \citet{Zhu0812,Mummery0523,Mummery0424,Mummery0624,Mummery0824} we consider only accreting BHs
that do not develop large-scale magnetospheres. It's not clear if these structures arrest or stop accretion (see e.g. \citealt{Nalewajko0724})
but they certainly strongly modify accretion's character.

\section{Disc accretion onto a black hole}
\label{sec:properties}

A rotating accretion flow onto a black hole cannot be fully described by the \citet{SS73} solution which by assumption neglects
radial pressure gradients and explicitly requires the radial flow velocity to be much less than the isothermal sound velocity: $\ll c_s$. Since at the horizon the radial velocity
is equal to the velocity of light this approximation must somewhere break down since the flow will have to pass through a sonic point.
Also the ``general-relativistic'' version of the $\alpha$-disc solution by \citet{NT73} is not adequate for the description of the motion of matter
near the BH event horizon because it requires more than just multiplying the \citet{SS73} solution by relativistic corrections.
To account fully and correctly for the properties of a BH-accreting {\sl fluid} such a description must include the full general-relativistic space-time structure.

The horizon is a null 3D surface (hypersurface) which has consequences that can be described only in the full Einstein's 
relativistic theory of gravitation. When interpreting these consequences one should avoid confusion and misunderstandings. For example, the presence of the horizon requires that matter crosses it at the ``speed of light''
(see e.g. \citealt{Abramowicz1196}). It does not mean, however, that matter's four velocity $u^i$ becomes a null vector and that $u^iu_i=0$. This is fundamentally impossible and one still has $u^iu_i= -1$.

``Crossing with the speed of light'' means that one can define a suitable {\sl radial velocity} which in a suitable reference frame will be equal to $1$ ($c$) at $r=r_H$. Explaining this requires a bit of general relativity that we will try to make as simple as possible. We will follow \citet{Abramowicz1196} but, for simplicity, assume that the BH is non-rotating, i.e. described by the Schwarzschild metric,
which in the equatorial plane is given by
\begin{equation}
    d s^2=-\tilde{\Delta} d t^2+ {r^2}d \varphi^2+\frac{1}{\tilde{\Delta}} d r^2,
\end{equation}
where $\tilde{\Delta}= 1 -2M/r$. At the horizon $\tilde{\Delta}=0$, i.e the horizon is at radius $r_H=2M$.

In a static spacetime the ``suitable'' reference frame will be that of static
observers, i.e. observers located at a fixed, constant distance $r$ from BH. Such observers exist
only for $r>r_H$ since at $r=r_H$ a static observer would have to move with the speed of light.

In the reference frame of a local static observer the four-velocity can be written as
\begin{equation}
\label{eq:fourv}
    u^i=\gamma\left(n^i+v^{(\varphi)} \tau^i{ }_{(\varphi)}+v^{(r)} \tau^i{ }_{(r)}\right),
\end{equation}
where $n^i$ is the four-velocity of the static observer, $\tau^i{ }_{(\varphi)}$ and $\tau^i{ }_{(r)}$ are unit vectors in the coordinate $\varphi$ and $r$ directions respectively, while
\begin{equation}
    \gamma=\frac{1}{\sqrt{1-\left(v^{(\varphi)}\right)^2-\left(v^{(r)}\right)^2}}.
\end{equation}
The angular velocity with respect to a static observer is defined by
\begin{equation}
  \Omega=\frac{u^{\varphi}}{u^t}= \frac{\tilde{\Delta}^{1/2}}{r}v^{(\varphi)},
\end{equation}
or, defining  a gyration radius $\tilde R=r\tilde\Delta^{-1/2}$, one can write
the azimuthal component of the physical velocity as $v^{(\varphi)}=\tilde R\Omega$.

The radial component of the four-velocity can be written in a ``special-relativistic'' form
\begin{equation}
    u^r =g_{rr}^{-1/2}\gamma v^{(r)}= \tilde\Delta^{1/2}\frac{V}{\sqrt{1-V^2}},
\end{equation}
which defines a physical radial velocity $V$. Since one can easily show that
\begin{equation}
    \gamma^2=\left(\frac{1}{1-{\Omega}^2 \tilde{R}^2}\right)\left(\frac{1}{1-V^2}\right)\equiv \gamma_\varphi^2\gamma_r^2
\end{equation}
the radial velocity $V$ can be written as
\begin{equation}
    V=\frac{v^{(r)}}{\sqrt{1-\left(v^{(\varphi)}\right)^2}}=\frac{v^{(r)}}{\sqrt{1-\tilde{R}^2 \Omega^2}}=\gamma_\varphi v^{(r)},
\end{equation}
thus $V$ is the radial velocity measured by an observer rotating with fluid at a fixed $r$. 

This radial velocity can be also written in the form of $V=1/\left[1+\tilde\Delta /\left( u^r u^r\right)\right]^{1/2}$,
which shows that at the horizon, where $\tilde\Delta=0$, one has $V=1$ {\sl independent} of the value of $u^i$.

\section{Equations}
\label{sec:equations}

The general-relativistic equations describing accretion onto a BH are
\begin{equation}
\label{eq:equation1}
    \nabla_i(\rho u^i)=0
\end{equation}
and \begin{equation}
\label{eq:equation2}    
\nabla_iT^i_k =0,
\end{equation}
where $\rho$ is the rest-mass density and $T^i_k$ is the stress-energy tensor.

Below we will present the simplest form of the SDA equations  since our aim is to identify the basic features that
are independent of the complexities we are here neglecting. One should add that the SDA is also applicable
to very low luminosity accretion flows, i.e. ADAFs \citep{Lasota0194,Abramowicz1196}.

The SDA \citep{Abramowicz0988,Sadowski0809} is a framework that allows taking into account
both the general-relativistic aspects of the accretion flow and the effects such as radial pressure gradients (negligible
in the main disc's body but crucial near its inner edge) necessarily neglected in the \citet{SS73} approach. The SDA \citep{Abramowicz0988}
was devised to describe accretion flows with luminosities close to the Eddington luminosity $L_{\rm Edd}={4\pi GM m_p c}/{\sigma_T}$ ($m_p$ is the proton mass, and $\sigma_T$ the Thomson scattering cross-section), 
but it also describes well the disc inner structure at lower luminosities, in particular flows at few tenths of $L_{\rm Edd}$ such as numerically described by \citet{Zhu0812}. At present the SDA paradigm is the sole framework that take into account
the fundamental property of fluid accretion onto a BH, namely that although passing through the sonic point is necessary for the existence of a
steady state accretion, it does not in itself guarantee the existence of a physically sound global solution.

In the following we will use the basic form of the stress-energy tensor : $T^{ik}= \rho u^iu^k + pg^{ik}$ and vertically integrated quantities, such surface density $\Sigma=\int_{-\infty}^{+\infty} \rho dz$ etc.
Then Eq. (\ref{eq:equation1})\jp{,} the equation of

{-- Mass conservation}\\
can be written as

\begin{equation}
\label{eq:masscons}
    \dot{M}=-2 \pi r \tilde\Delta^{1 / 2}  \Sigma\frac{V}{\sqrt{1-V^2}}.
\end{equation}

Various components of Eq. (\ref{eq:equation2}) provide the other conservation equations.

\noindent {-- Momentum conservation}

\begin{equation}
\label{eq:momcons}
    \frac{V}{1-V^2} \frac{d V}{d r}= \frac{\mathcal{A}}{r}-\frac{1}{\Sigma} \frac{d P}{d r},
\end{equation}
where
\begin{equation}
    \mathcal{A}=-\frac{M r}{\tilde\Delta \Omega_K^2} \frac{\left(\Omega-\Omega_K\right)^2}{1-{\Omega}^2 \tilde{R}^2},
\end{equation}
$P$ the vertically integrated total pressure and $\Omega_K=\sqrt{{M}/{r^3}}$ is the Keplerian angular velocity.\\

\noindent {-- Energy conservation}

\begin{equation}
\label{eq:energcons}
    F^{\rm adv}= F^+ - F^-,
\end{equation}
where the advective cooling is
\begin{equation}
  F^{\rm adv}= \frac{\tilde\Delta}{r\sqrt{1-V^2}} T \frac{dS}{dr}= -\frac{\dot M}{2\pi r}T \frac{dS}{dr},
\end{equation}
where $S$ is the entropy and $T$ the midplane temperature,
and the ``viscous'' heating
\begin{equation}
    F^+ = \Sigma \nu\gamma_{\varphi}^4 r^2\left(\frac{d\Omega}{dr}\right)^2,
\end{equation}
whereas the optically thick cooling has the form
\begin{equation}
    F^- = \frac{16 \sigma T^4}{3\tau},
\end{equation}
where $\sigma$ is the Stefan-Boltzmann constant and
$\tau$ the total, vertical optical depth.

\noindent {-- Angular momentum conservation}

\begin{equation}
\label{eq:amcons}
\frac{d }{d r}\left(r \tilde\Delta^{1/2}\frac{V}{\sqrt{1-V^2}}\Sigma \mathcal{L}\right) + \frac{d}{dr}\mathfrak{T} =0,
\end{equation}
where $\mathcal{L}=u_\varphi$ is the specific (per unit rest mass) angular momentum (in the non-relativistic description called $\ell$, see Eq. \ref{eq:bp}) and $\mathfrak{T}$ is the ``viscous'' torque driving accretion. In ionised accretion disc the turbulent anomalous viscosity is generated by the magnetorotational instability
(MRI; \citealt{Balbus0791,Balbus0198}).

\section{The global transonic solution}
\label{sec:glob}

In this section we will present the fundamental properties of BH accretion flows
forced on by the constraints imposed by the general relativity theory.

\subsection{Torque at the horizon}

Using Eq. (\ref{eq:masscons}), Eq. (\ref{eq:amcons}) can be integrated to
\begin{equation}
 - \dot M  \mathcal{L} + \mathfrak{T} = const.
\end{equation}
At the horizon $r=r_H$, the torque must be equal to zero $\mathfrak{T}(r_H)=0$,
hence the equation of angular momentum conservation can be written as
\begin{equation}
 \dot M  \left(\mathcal{L} - \mathcal{L}_H\right)=2\pi \mathfrak{T}.
\end{equation}
The value of the integration constant $\mathcal{L}_H$, i.e. of the angular momentum at the horizon is neither arbitrary nor known a priori and must be calculated through solving the other conservation equations (\citealt{Abramowicz0189} and references therein). Although it has been often assumed that the torque vanishes at some chosen location of the accretion flow (most often at the ISCO) only the surface of a BH is the place where it must vanish since no flux of angular momentum, or anything else, can cross this surface outwards. That is why the only way to slow down BH rotation is to feed it with negative angular momentum \citep[see e.g.,][]{Lasota0114}.

For a quarter of a century there have been unending discussions about the locus of the vanishing torque and its value at the ISCO. The discussions started with the ``KGA''\footnote{Acronym used by \citet{Paczynski0400}.} \citep{Krolik0499,Gammie0999,Agol0100} papers and have continued until now despite the clarifying and, one could have hoped, sobering article by \citet{Paczynski0400}. Unfortunately this clear and short preprint\footnote{This fundamental paper has been rejected from the Astrophysical Journal Letters because ``the referee could not be convinced'' \citep{Afshordi0703}.} has been ignored or misunderstood. Most of its readers probably missed its main point: the value of the torque everywhere in the accretion flow is determined by the laws of physics, or more specifically by the unique transonic solution corresponding to the value of the specific angular momentum at the BH surface where the torque necessarily vanishes. In other words the value of the torque at a given place in the accretion
flow cannot be assumed arbitrarily because it is determined by the value of the flow's angular momentum at the BH surface.

In general it is assumed that the torque is due to a turbulent stress acting as
viscosity:
\begin{equation}
    \tau_{\varphi r}=\overline{\rho v_{\varphi}v_r} - \overline{B_{\varphi} B_r} = - \Sigma \nu r \frac{d\Omega(r)}{dr},
\end{equation}
where $\nu$ is the kinematic viscosity coefficient and $v_r$, $v_{\varphi}$, $B_r$, $B_{\varphi}$ are respectively the 
velocity and magnetic-field radial and azimuthal components and $\Omega$ the angular velocity.

Including the general-relativistic effects one obtains for the ``viscous'' torque
\begin{equation}
    \mathfrak{T} = -\Sigma \nu \tilde\Delta^{1/2}r^3\gamma_\varphi\frac{d\Omega(r)}{dr},
\end{equation}
in case of a non-rotating BH.

Despite the discovery of the MRI origin of the turbulence in ionised quasi-Keplerian accretion discs,
to describe accretion discs one still often uses the $\alpha$ viscosity prescription in its various forms.

\citet{Mummery0424} argue that near the ISCO the ``classical'' viscous descriptions of turbulent fluids 
is inadequate and propose a different description of this part of the accretion flow (see also \citealt{Mummery1022,Mummery0523}). 
Then \citet{Mummery0624}
use this new approach to interpret the numerical results of \citet{Zhu0812}. This is somehow surprising
because the te GRRMHD simulations in question solve the standard classical equations, using values
of $\alpha$ to parametrise their solutions. We will come back tho this problem later after discussing
the fundamental properties of the solution of Eqs. (\ref{eq:amcons}), (\ref{eq:masscons}), (\ref{eq:momcons}) and (\ref{eq:energcons}).

\subsection{The sonic point}

Integration of the angular-momentum conservation equation provides a constant $\mathcal{L}_H$. To determine its value
one has to integrate the remaining conservation equations.

Assuming for simplicity that the BH accretes a perfect gas
\begin{equation}
    P =\frac{\mathfrak{R}}{\mu}\rho T,
\end{equation}
the remaining BH accretion-flow conservation equations can be solved for $dV/dr$ and $d\Sigma/dr$:
\begin{equation}
    \frac{d\ln V}{d\ln r}=\frac{N_1}{D}(1-V^2),
\end{equation}
and 
\begin{equation}
    \frac{d\ln \Sigma}{d\ln r}=\frac{N_2}{D},
\end{equation}
with
\begin{equation}
    {D}= \mathfrak{a}_s^2- V^2
\end{equation}
where $a_s^2=\Gamma_1 c_s^2$ is the adiabatic speed of sound and $\Gamma_1$ is the ratio of specific heats.
The numerators  on the right hand sides of the conservation equations are respectively
\begin{equation}
    N_1=-\mathcal{A}-\frac{r-M}{r\tilde\Delta} \mathfrak{a}_s^2-\frac{\left(\Gamma_1-1\right) 2 \pi r^2}{\dot M}F^{\rm adv}
\end{equation}
and
\begin{equation}
    N_2=\frac{r\tilde\Delta N_1 -D(r-M)}{r\tilde\Delta}.
\end{equation}
The sonic point $r_s$ where $V(r_s) =\mathfrak{a}_s(r_s)$ is a singular point of the equations unless
$N_1(r_s)=N_2(r_s)=0$. The latter equations determine the topology of the sonic point \citep{Abramowicz0189,Afshordi0703}.

\subsection{The transonic solution}

\begin{figure}
\hspace*{-0.4cm}
\includegraphics[width=0.5\textwidth]{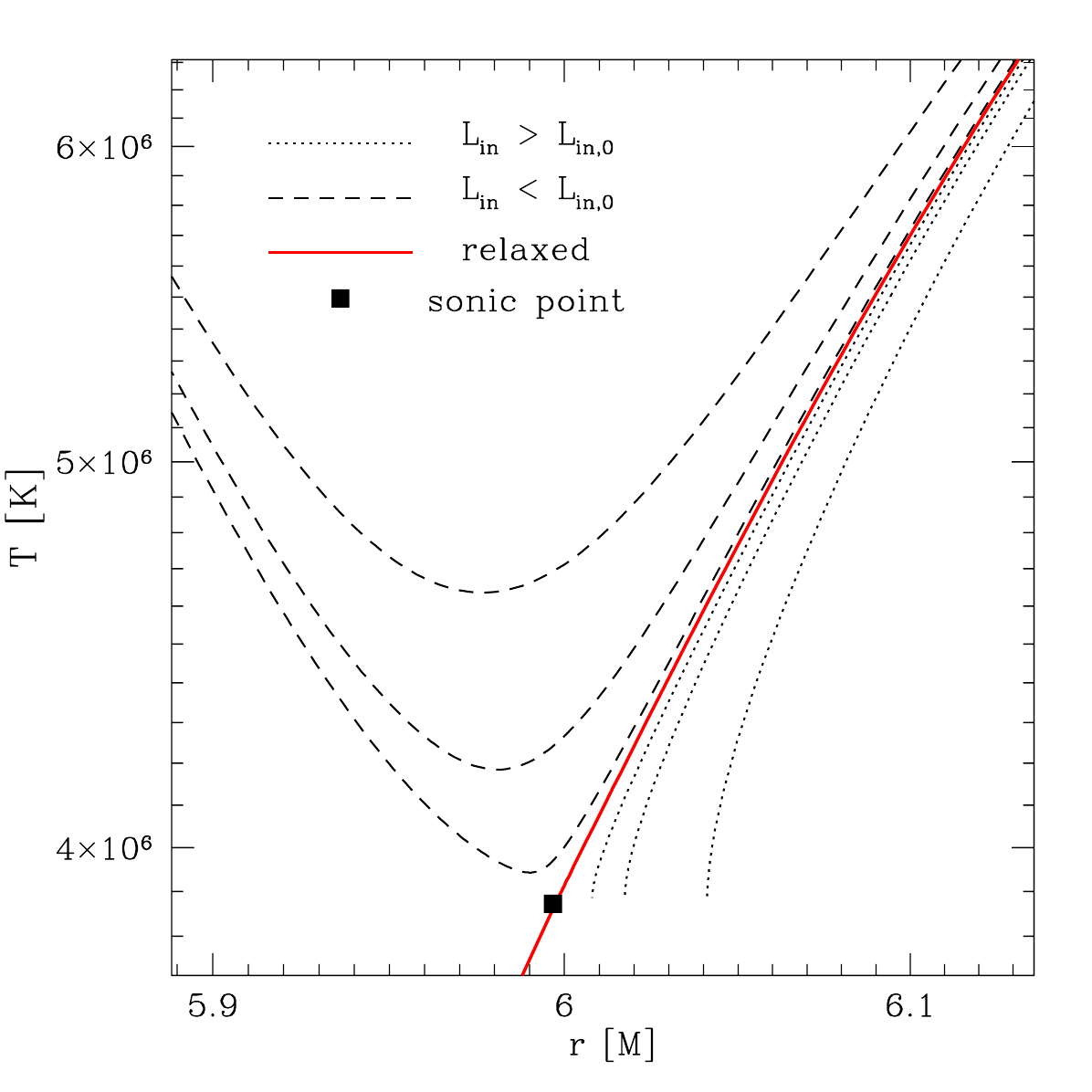}
\caption{Temperature profiles in the vicinity of the sonic point for several iterations
leading to the trial model used as initial condition in the relaxation procedure. Solutions with too high value of $L_{\rm in}$ terminate before reaching
the sonic point while solutions with too low $L_{\rm in}$ go through the sonic point radius
but follow an improper branch. The relaxed solution and the location of the sonic point
are also presented. Models calculated for a non-spinning BH with $\dot M = 0.1 \dot M_{\rm Edd}$. From \citet{Sadowski0811}.
}
\label{fig:sonic}
\end{figure}
Every solution describing physically sound accretion onto a BH {\sl must be transonic}. 
Far from the BH accretion forms a subsonic flow in which $V \ll \mathfrak{a}_s$, a
\citet{SS73} type of disc, or a subsonic quasi-radial inflow. Since at $r_H$ the radial velocity
must be $V=1$ it follows that, if the accreting matter's equation of state satisfies the causality
constraint $\mathfrak{a}_s < 1$, the flow must pass through a sonic point located outside the BH surface.
Since $V=1$ at $r=r_H$ and $V\ll \mathfrak{a}_s$ when $r \rightarrow \infty$ are a fundamental property
of accretion onto BHs, the transonicity of the solution describing this process is its inescapable
property. It is therefore rather surprising that the notion of the sonic point fails to appear in the papers about the torque at ISCO
\citep[e.g., in][]{Zhu0812,Mummery0424,Mummery0624,Mummery0824}, even if in some of them the supersonicity of the inner accretion flow is mentioned.

The sonic point is a critical point of the conservation equations (\citealt{Abramowicz0189} and references therein). 
For a solution of the equations describing accretion onto a BH to be regular and globally acceptable  
it is necessary, although not sufficient, for the critical point to be of the saddle or nodal
type. However, the sonic point/local regularity conditions, together with
the global topological constraints, cannot be met for all the
choices of boundary conditions and parameters describing the
material properties (equation of state, opacity, viscosity) of the accretion flow. This
creates forbidden regions in the parameter space of the problem
and means that not all astrophysically acceptable boundary
conditions can lead to regular stationary accretion flows \citep{Abramowicz0189}.

In practice this means that neither the value of $\mathcal{L}_H$ nor the location of the sonic point $r=r_s$ are known a priori and must be found by solving the equations describing the flow. For a given set of outer boundary conditions (usually corresponding to
the stationary \citealt{SS73} solution in its \citealt{NT73} version), given the black-hole mass $M$ and spin $a_{\rm BH}$ and assuming an
accretion rate $\dot M$, it is the value of $\mathcal{L}_H$ that
determines the shape of the solution, i.e. the passage through the sonic point. Figure \ref{fig:sonic}
illustrates an example of the attempts at solving accretion-onto-BH equations by the relaxation technique \citep{Sadowski0811} 
that starts from a trial solution. The figure show the sensitivity of the solution to the choice of $\mathcal{L}_H$ (here called 
$L_{\rm in}$): even a small deviation from the proper value of $\mathcal{L}_H$ makes the function supposed to be a solution of the equations to miss the sonic point. This means that the value of the angular momentum at the surface of the BH fully determines the
value of the accretion torque {\sl everywhere} in the flow. In other words, modifying the value the torque e.g. at the ISCO, without
checking if it is consistent with $\mathcal{L}_H$ might, and in most cases will, result in a function not corresponding to a transonic
solution. It is mainly the general-relativistic constraints that guide the accretion flow through the space-time.

It should be stressed that the example of Figure \ref{fig:sonic} corresponds to an accretion flow with $L/L_{\rm Edd}=0.1$, i.e.
a luminosity {\sl lower} than the luminosities ($L/L_{\rm Edd}=0.32 - 0.71$) considered by \citet{Zhu0812}. It makes sense therefore to compare the solutions of \citet{Sadowski0809,Sadowski0811} with
results of \citet{Zhu0812} simulations.

\section{SDA vs GRRMHD and geodesics solutions}
\label{sec:versus}

\subsection{A transonic solution in the SDA}

To illustrate the properties of the BH-accretion global transonic solution we will use the
the results of \citet{Sadowski0809,Sadowski0811}.
Although this work uses the slim disc set of equations, contrary to a popular misconception
not all solutions of these equations correspond to an advection dominated flow. For example, when
$L=0.1 \Ledd$, the disc aspect ratio $H/r\approx L/\Ledd=0.1$ (the flow is then radiation-pressure
dominated) so that $F^{\rm adv}\approx (H/r)^2 F^+\approx 10^{-2}F^+$ and in this case advection plays a minor
role in the energy balance. Slim discs, i.e. advection dominated flows appear only at $L \gtrsim \Ledd$\footnote{A luminosity
at which appear also outflows (see \citealt{SS73})}. But the SDA allows one to take into account radial pressure
gradients and departures from Kepler's angular momentum distribution -- elements necessary for the
description of a transonic flow -- for much lower luminosities.
\begin{figure}
    \hspace*{-0.4cm}
    \includegraphics[width=0.99\linewidth]{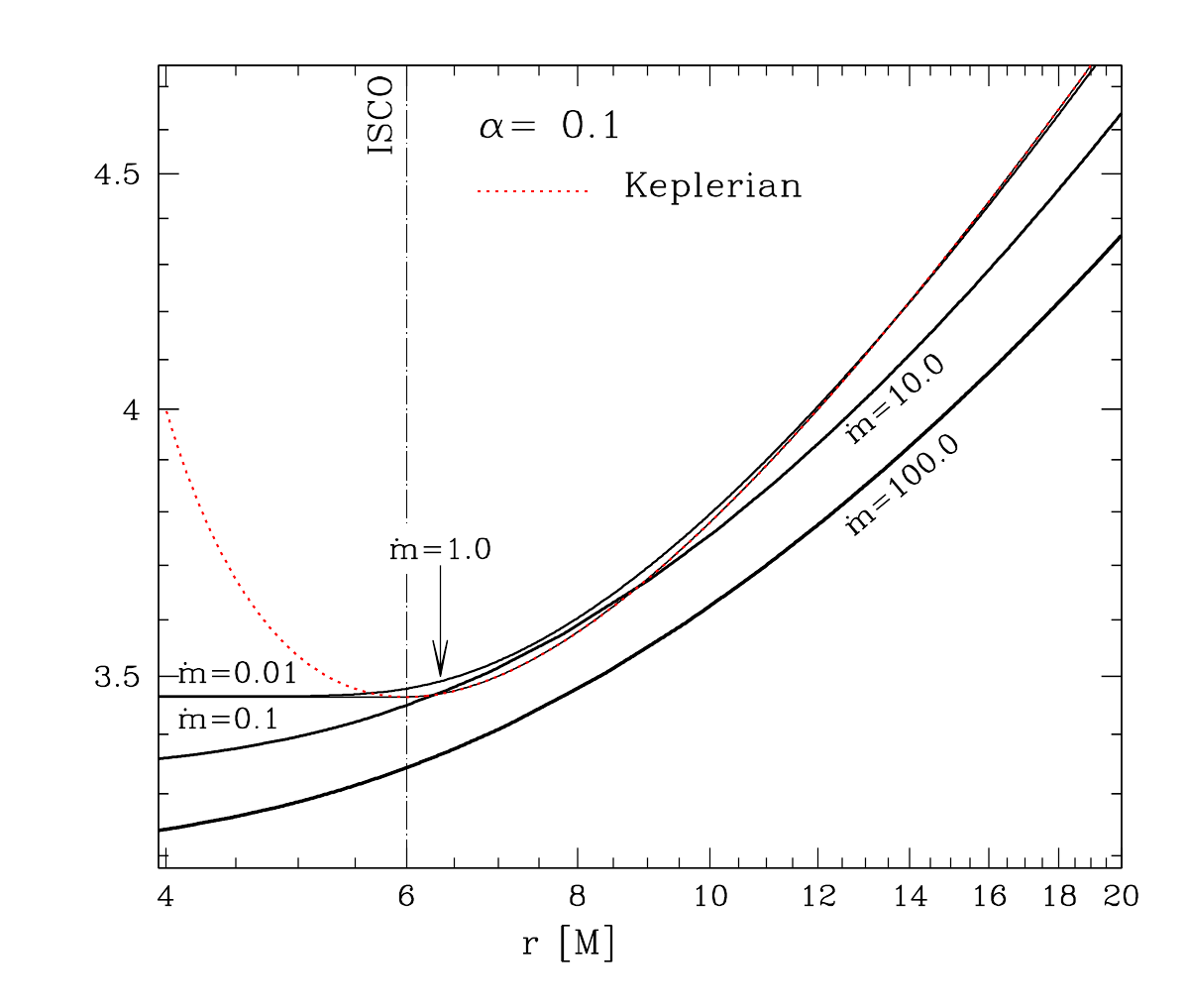}
    \caption{Profiles of the disk angular momentum $\mathcal{L}$ for  $\alpha = 0.1$
for accretion rates corresponding to luminosities $L/\Ledd = 0.01$, $0.1$, $1.0$, $10.0$ and $100.0$ (see the text for the explanation of the units). The spin of the BH is $a_{\rm BH} = 0$. From \citet{Sadowski0811}.}
    \label{fig:angmom}
\end{figure}
Figure \ref{fig:angmom} shows the distribution of the angular momentum of a transonic solution for five values
of the accretion rate, assuming $\alpha=0.1$\footnote{Since the definition of the Eddington accretion rate
depends on the assumed accretion efficiency (in S\c adowski's papers assumed to be 1/16), we prefer to use luminosities in
Eddington units instead of accretion rates $\dot m=\dot M/\dot M_{\rm Edd}$.}. One can see that the torque at ISCO becomes noticable
only for $L > 0.1\Ledd$ but even for $L =\Ledd$ is rather small. 

\begin{figure}
    \hspace*{-0.4cm}
    \includegraphics[width=0.99\linewidth]{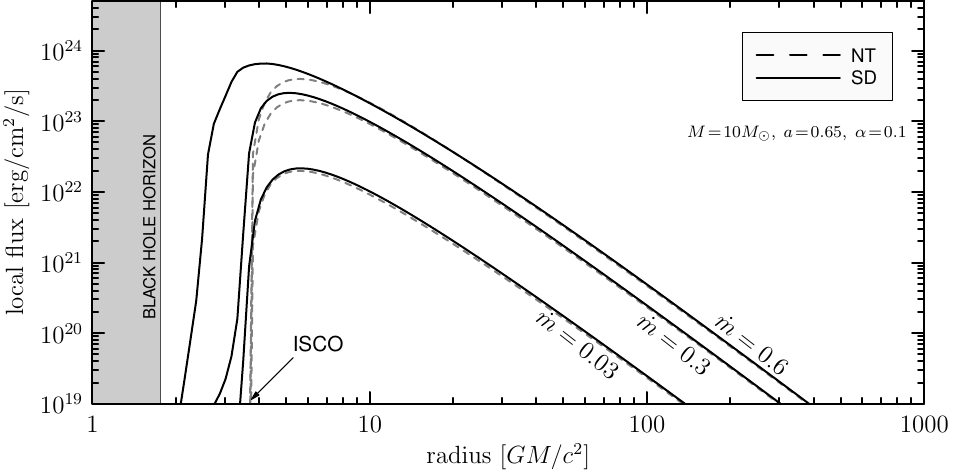}
    \caption{Flux profiles emitted from the surface of transonic disc (solid) and according to the \citet{NT73} model (dashed lines) for BH spin $a_{\rm BH}=0.65$. Fluxes are plotted for three different mass accretion rates $L/\Ledd = 0.03$, $0.3$ and $0.6$ . From \citet{Sadowski0811}.}
    \label{fig:fluxes}
\end{figure}

In Fig. \ref{fig:fluxes} we show fluxes emitted by
transonic discs calculated by \citet{Sadowski0811} compared to emission  obtained from the \citet{NT73} model in which the torque
at ISCO vanishes. We see that in a transonic disc a significant flux from the plunging region is emitted only for $L=\Ledd$, even if some tiny
amount of radiation is emitted for this region even for $L= 0.03\Ledd$.

This is to be compared with model C in Fig. 1 of \citet{Zhu0812} which corresponds to a similar configuration with 
$\dot m=0.3$ in Fig. \ref{fig:fluxes}. One can see that according to the GRRMHD simulations the flux from the
plunging region is roughly an order of magnitude larger than that obtained from the transonic disc. As claimed
by \citet{Zhu0812}, however, the main difference is that according to the GRRMHD calculation what is emitted there
is the energy produced by the viscous torque while in the case of the transonic disc the radiation from the plunging
region is emitted by matter advected from larger radii \citep{Sadowski0809,Sadowski0811,Abramowicz1010}.
We notice that it is not clear why in GRRMHD, at luminosities close to Eddington's, the advected flux should be negligible.

The value of the radiation flux in the plunging region and its origin, in the case of a the transonic disc is consistent
with the radial angular-momentum profile which for moderate values of $\alpha$ and sub-Eddington luminosities is almost flat 
between the BH surface and the ISCO indicating that between these two loci the torque is negligible. 

In a separate paper (Abramowicz \& Lasota 2024, in preparation) we show how the \citet{SS73} analytic formula for the flux $F(r)$ 
agrees with the ideas discussed here.

\subsection{Are solutions resulting from GRRMHD simulations globally transonic ?}

In the case of GRRMHD simulations it is not clear that they correspond to global transonic
solutions for which the torque vanishes at the horizon. The words ``sound'' or ``sonic'' do not appear\footnote{The word ``sound''
is used once in the sense ``based on valid reason''.}
in the article of \citet{Zhu0812}. 

Recently \citet{Mummery0424,Mummery0624} presented analytical solutions supposed to represent trans-ISCO accretion
flows and show that they represent faithfully the GRRMHD simulations of \citet{Zhu0812}. One can hope therefore that these solutions could provide information about
the true nature of the GRRMHD simulations.

\citet{Mummery1022,Mummery0523,Mummery0624} aim at describing the region  around the ISCO of the
accretion flow onto a BH. They describe the ISCO as (semi)absorbing limit for the flow
particles and define a parameter $\delj$ corresponding to the inner boundary condition for the stress at the ISCO. The parameter $\delj$ is supposed to correspond to the fraction of its ISCO angular momentum an accreting fluid element is able to send back from the plunging region to the disc's main body. 
It is the infra-ISCO stress ``excess'' compared to the case of zero-stress at the ISCO. 

\citet{Mummery0424} solve their equations for five free parameters: $M_{\rm BH}$, $a_{\rm BH}$, $\dot M$,
$\alpha$ and $\delta_{\mathcal{J}}$ (we will call this scheme MISCO). Here $\delta_{\mathcal{J}}$ plays the role of $\mathcal{L}_H$
in the SDA framework. However, $\mathcal{L}_H$ is not and cannot be an independent free
parameter \citep{Abramowicz0189} as explained above. This is of course true also of $\delta_{\mathcal{J}}$ since basically
the same system of equations should be solved in both cases, since its relevant properties do not depend on
the definition of the torque. 
From Eq. (\ref{eq:bp}) one obtains
\begin{equation}
    \delta_{\mathcal{J}} = 1 - \frac{\mathcal{L}_H}{\mathcal{L}\left(r_{\rm ISCO}\right)},
\label{eq:bp2}
\end{equation}
but since Eq. ({\ref{eq:bp2}}) has been derived from $V(r_{\rm ISCO})\approx V(r_s)=\mathfrak{a}_s$,
the value of $\mathcal{L}(r_{\rm ISCO})$ is {\sl fixed} by the global solution passing through the critical
point. As explained above, $\mathcal{L}_H$ is not a free parameter but has to be obtained by
guiding the global solution through the sonic point, the same must be true of $\delta_{\mathcal{J}}$.
Hence using $\delta_{\mathcal{J}}$ as an independent  {\sl free} parameter, corresponding to the value of the stress at ISCO
\citep{Mummery0624}, might produce non-solution \citep{Abramowicz0189,Afshordi0703}.

One should also point out that, in principle, the MISCO scheme cannot
provide a global solution because their analytical models of trans-ISCO flows involve piecewise joining 
of intra- and extra-ISCO flows.

Keeping this in mind we will have a closer look at the correspondence between the results of the GRRMHD simulations
and MISCO. For example for the model C mentioned above, \citet{Mummery0624} found $\delj = 0.0085$, which corresponds
to $\mathcal{L}\left(r_{\rm ISCO}\right)=10086 \mathcal{L}_H$, so the stress between the ISCO and the BH horizon
is rather small. But the MISCO has some unusual properties. It fixes the velocity at the ISCO as
\begin{equation}
    U^r\left(r_{\rm ISCO}\right)= -\alpha^{1/2}c_s\left(r_{\rm ISCO}\right),
\end{equation}
where $U^r$ is the radial component of the four velocity in the (presumably) stationary (ZAMO) reference frame and
$\alpha$ is the standard viscosity parameter. Then a trans-ISCO velocity profile is defined as \citep{Mummery0424}
\begin{equation}
    U^r\left(r\right)= U^r_{\rm NT}\left({r}\right) - \alpha^{1/2}c_s\left(r_{\rm ISCO}\right)\mathcal{I}(r),
    \label{eq;visco}
\end{equation}
where $U^r_{\rm NT}$ is the radial velocity in the \citet{NT73} model and $\mathcal{I}$ an interpolation function.
Eq. (\ref{eq;visco}) is not a solution of any equation but is removing the cusp in the physical parameters when one
assumes that matter moves on geodesics inside and outside the ISCO \citep{Mummery0424}. The resulting radial velocity
profiles for $r \leq r_{\rm ISCO}$ are independent of $\delj$, i.e.. independent on the ISCO stress, while other
parameters such as the effective and midplane temperatures and density (but not the surface density linked to the radial
velocity through the mass conservation equation) vary by more than one order of magnitude when $\delj$ is varied
by four orders of magnitude. 
\begin{figure}
    \centering
    \includegraphics[width=0.99\linewidth]{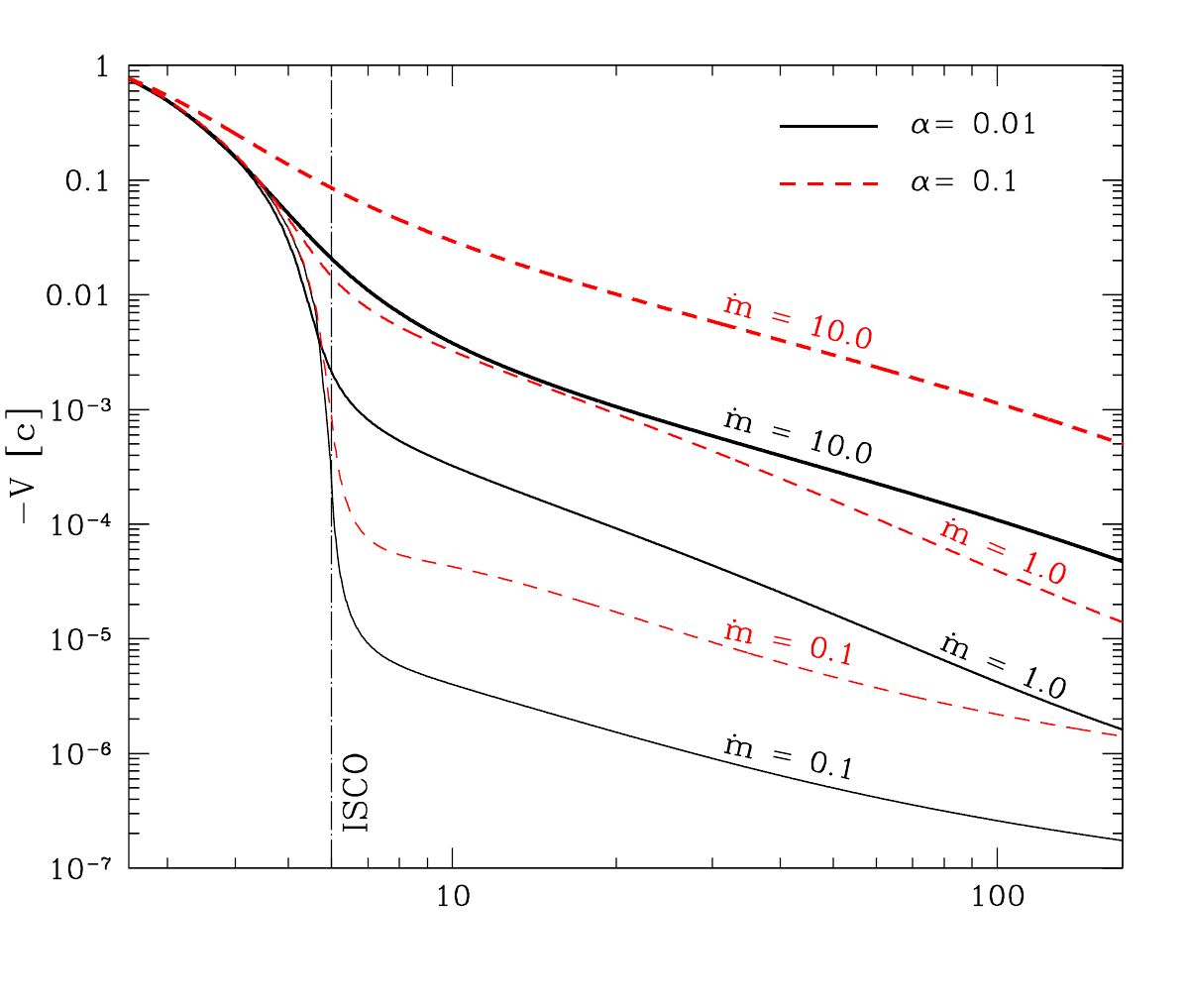}
    \caption{Profiles of the radial velocity for global transonic disc solutions for various accretion
rates. The BH has $M_{\rm BH}=10\msun$ and $a_{\rm bh}=0$. Solid black
lines present solutions for $\alpha=0.01$ while red dashed lines for $\alpha=0.1$. From \citet{Sadowski0811}.}
    \label{fig:vrs}
\end{figure}
It is not explained why the radial velocity is totally independent of the stress. The information that ``the trans-ISCO fluctuation
velocity turns out to be only weakly dependent on the ISCO stress in this new formalism.''\citep{Mummery0424}
is not very helpful. Figure 8 in \citep{Mummery0424} looks similar to our Fig. \ref{fig:vrs} but in the latter
each velocity profile correspond to a one specific value of $\mathcal{L}_H$, i.e. to one specific value of $\delj$.
At sub-Eddington accretion rates these profiles are close to each other since then the intra-ISCO stress is small (but not zero). 
The explanation of the weird properties of the ``new formalism'' is rather simple: particles of the accretion flow move
on geodesics with the same initial conditions but electromagnetic radiation is calculated using a value of the assumed stress $\delta_j$,
unrelated to the particle motion since energy and angular momentum in geodesic motion are
conserved and therefore matter not losing energy cannot radiate. In other words the ``new formalism''
is physically inconsistent.

\citet{Mummery0624} don't use the refined version of the model and assume that in the disc (supra-ISCO) matter
follows circular Keplerian orbits and infra-ISCO it falls in along geodesics calculated by \citet{Mummery1022}.
Unfortunately, as pointed above, this is one of the weak points of MISCO: the accretion flow fluid lines do not coincide with geodesics
and the plunging region does not originate at the ISCO. At low accretion rates the fluid lines in the body of the
disc are not geodesics, only their circular component is almost, but not exactly Keplerian. The full motion cannot be geodesic,
after all the radial velocity is $\sim \nu/r$. Also below the ISCO matter is not in free fall.
The plunging region begins at the so-called
potential spout edge $r_{\rm pot}$, where the specific angular momentum of the disc crosses the Keplerian profile from above
(see Figure \ref{fig:roche} and sect. \ref{sec:discussion} and \citealt{Abramowicz1010}. The spout edge is always below the ISCO, $r_{\rm pot}< r_{\rm ISCO}$. Even for $L/\Ledd=0.1$ 
$r_{\rm pot}= 5.9 M$, for $a_{\rm BH}=0$ \citep{Sadowski0811}. The velocity gradient and the pressure gradient terms in Eq. (\ref{eq:momcons})
are small but not negligible, especially near the disc's inner edges \citep{Abramowicz1010}. 
This is one of the reasons why equations describing accretion flows onto BH have critical points and their
solution must pass trough them. That is why MISCO is not very helpful in understanding the mathematical and
physical nature of the GRRMHD-simulation solutions. 

The GRRMHD simulations solve the usual $\nabla_i T^i_k = 0$ equations, i.e. they solve for
velocities, densities, temperatures etc. and it is not clear how their solutions could
represent effects of velocity fluctuations close to the ISCO, developing a non-zero directional bias
as found e.g. in \cite{Mummery0424}.

On the other hand \citet{Straub0911} used the solution of slim-disc equations to fit the X-ray spectra
of LMC X-3 so that using the SDA paradigm seems to be the proper framework for understanding
the GRRMHD simulations of BH accretion flows.

\subsection{Observational determination of $\delj$ ?}
\label{sec:1820}

\citet{Mummery0624} claim that by taking into account the plunging-region emission they removed the need of adding ad hoc components when fitting the X-ray spectra of X-ray binaries. Below we will show that this statement is unsubstantiated but first we have to notice that
for example assuming that the the disc is covered by a warm Comptonised layer \citep{Belczynski24,Belczynski1024,Zdziarski0224b}
is no more ad hoc, than assuming in TLUSTY that energy dissipation follows the vertical density profile. i.e. assuming that 
dissipation is maximal at the mid-plane \citep{Done2008}. This assumption is known to be not necessary \citep{Hubeny1098}, contradicting 
both MRI simulations \citep{Blaes0914} and observations \citep{Hubeny0621}. It is widely used in the disc-fitting procedure but
it does not make it credible.

\citet{Mummery0624} estimate the value of $\delj$ by fitting the MISCO model
to the X-ray spectrum of the BH X-ray transient MAXI J1820+070. They do it in a rather novel way. First, 
they fix the black hole spin and source-observer inclination to values inferred from the fit to the hard-state X-ray spectrum, 
that are $a_{\rm BH}=0.2$ and $i=35^{\circ}$ (\citealt{Buisson1119,Dias0424}, interestingly most authors find
$i \gtrsim 60^{\circ}$; see Table 3 in \citealt{Thomas0122}), then they fit the soft-state spectrum of this source with the
MISCO model obtaining $\dot M = 0.86 \pm 0.02\,\Ledd/c^2$ and $\delj=0.0385 \pm 0.0014$.
\citet{Mummery0624} claim that they demonstrated that the detection of intra-ISCO emission forces the MAXI J1820+070 BH spin 
be low: $a_{\rm BH}\leq 0.52$ ($99.9 \% $ confidence level). This has not prevented \citet{Banerjee0424} fitting the soft-state spectrum of
MAXI J1820+070 with $a_{\rm BH}=0.77$ and $i=64^{\circ}$. It does mean that the latter determination is correct but is just another demonstration
that formal solutions for BH spin from spectra with their reported extremely small error bars or very
definitive values are model-dependent and are not the only possible result of spectral fitting \citep{Belczynski24,Belczynski1024}.
So much for the ``the first robust detection of intra-ISCO emission in the literature'' and the determination
of the parameter $\delj$.

\section{Discussion}
\label{sec:discussion}

There is no reason, physical or mathematical, for the stress to vanish at the ISCO.
It has been used as a convenient inner boundary condition by \citet{SS73}, but in
the general-relativistic case of the \citet{NT73} model it leads to nonphysical
consequences, such as infinite radial velocity as noticed long time ago by \citet{Stoeger1276}.
\begin{figure}
    \includegraphics[scale=0.73]{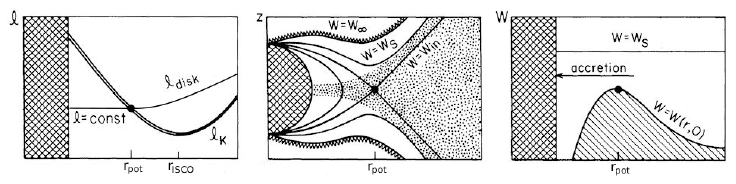}
    \caption{An illustration of the accretion-flow Roche lobe overflow. The leftmost panel schematically presents disk angular momentum profile and its relation to the Keplerian distribution. The middle panel shows the equipotential surfaces. The dotted region denotes the volume filled with accreting fluid. The rightmost panel presents the potential barrier at the equatorial plane (z = 0) and the potential of the fluid (WS) overflowing the barrier. 
    (\url{http://www.scholarpedia.org/article/Accretion_discs}).
    }
    \label{fig:roche}
\end{figure}

The problem of the inner structure of accretion flows onto BH has been solved by \citet{Kozlowski0278}
who developed Paczy\'nski's idea of BH accretion through ``Roche-lobe overflow'' as shown in Fig. \ref{fig:roche}.
Paczy\'nski (unpublished) noticed that the equipotentials in the problem of accretion onto a BH
have a similar structure (with different 3D topology) to equipotentials in a binary system.
A description of fluid lines by geodesics is obviously incompatible with such representation of the BH accretion flow.
As mentioned above the ``L1'' point (spout) is close to, but not coincident with the ISCO. None of the various
inner edges of the accretion flow \citep{Krolik0702,Abramowicz1010} coincide with the ISCO but for low accretion
rates they are very close to this critical orbit.
In such a configuration the stress can vanish only on the BH surface as it has no reason to
do it in any other place in the accretion flow.

Of course this does not preclude the properties of the stress to be affected by the vicinity of the
ISCO, however, when the stress is due to inner physical processes in the flow the resulting
modifications must be consistent with the global flow properties. They must be calculated, they cannot be described by
a free parameter.

\section{Conclusions}
\label{sec:conclusions}

Astrophysical accretion onto a BH must be described by global solutions of the (M)HD equations in the
Kerr spacetime. In the stationary case these solutions (of the conservation equations) must be transonic
and their properties are determined by the value of the specific angular momentum at the BH surface
where the the inner-fluid torque must vanish. This angular-momentum value and the location of sonic point
cannot be known a priori because not for all their pairs a solution exists. The space-time structure
imposes inescapable constraints on the physically sound models of accretion onto BHs. 

At present only the SDA solutions are ‘topologically’ correct: they take into account
the critical points whose existence follows from the local and {\sl global}
properties of the space-time and the local description of physics.
This is the best semi-analytical description of BH accretion available at the moment.
It has been successfully applied to fitting X-ray spectra \citep{Straub0911}.
It certainly can be used to interpret results of numerical simulations.

Finally we enumerate some basic facts about accretion onto BHs.

\begin{itemize}
    \item Everywhere -- above and below the ISCO -- accretion discs are {\sl fully} described 
by the equations $\nabla_i T^i_k = 0$, but the specification of the
$T^i_k$ tensor is difficult -- the description of the stress is incomplete
analytically \citep{Balbus0899}. Hence the alpha prescription and, for example, the slim disc approximation.
    \item The description of the $T^i_k$ tensor is numerically ‘complete’ in the GRRMHD,
          It is based on partial differential equations that are local: they do not include non-local interactions
i.e. these which act on a finite distance.
    \item Modelling radiation of black hole accretion discs with
geodesic motion of matter is fundamentally inconsistent (nonphysical)
because energy and angular momentum in geodesic motion are
conserved and therefore matter does not lose energy and
cannot radiate. Assuming geodesic motion between the ISCO and the horizon is equivalent to assuming that the stress at the ISCO is the
same as at the horizon, i.e. zero. Thus it cannot be
a free parameter.
     \item Fluid current lines in accretion discs do not coincide with
geodesic orbits. However, above the ISCO, for low viscosities and low accretion
rates they almost coincide with circular
geodesics but are always sub-Keplerian. In these
circumstances the Shakura-Sunyaev model is an excellent description
of accretion discs. Above the ISCO it assumes Keplerian motion, zero stress at the ISCO and the geodesic free-fall
below this disc edge.
     \item Detection of plunging-region emission in spectra of X-ray binaries is as credible
     as BH spin determinations from fitting these spectra with various disc radiation models.
\end{itemize}

\begin{acknowledgement}
JPL thanks Andrzej Zdziarski for drawing his attention to the recent attempts to address the problem of the plunging-region emission
and both authors thank him for very helpful comments on an earlier version of our paper.
\end{acknowledgement}

\bibliographystyle{aa} 
\bibliography{plunge} 

\end{document}